\documentclass[aps,prd,longbibliography ,twocolumn,nofootinbib,10pt]{revtex4-2}
\usepackage{amsmath,amsfonts,amsthm,amssymb,mathtools}
\usepackage{hyperref,makecell,float,chngcntr}
\usepackage{xcolor,bm,braket, comment}

\usepackage{tikz}
\usepackage{mathtools}  
\usepackage{xfrac}

\usepackage{graphicx} 

\begin{document}

\title{A modified sine-Gordon theory with static multi-kinks}
\author{Chris Halcrow}
\email{chalcrow@kth.se}
\affiliation{Department of Physics, KTH-Royal Institute of Technology, SE-10691, Stockholm, Sweden}
\author{Renjan Rajan John}
\email{renjan@mgu.ac.in}
\affiliation{School of Pure and Applied Physics, Mahatma Gandhi University, Kottayam, Kerala 686 560, India}
\author{Anusree N}
\email{anusreekanakaraj@gmail.com}
\affiliation{Department of Physics, Sree Narayana College, Kannur, Kannur University, Kerala 670 007, India}
\date{\today}
 \begin{abstract}
    We construct a modified non-BPS sine-Gordon theory which supports stable static kinks of arbitrary topological degree $N$. We use this toy model to study problems which are interesting for higher-dimensional soliton theories supporting multi-solitons. We construct a 2-kink collective coordinate model and use it to generate scattering trajectories, which are compared to full-field dynamics. We find that the approximation works well, but starts to fail as radiation becomes more important due to our model becoming less BPS or when the initial kink velocities are large. We also construct the quantum 2-kink and calculate one-loop corrections to the 1- and 2-kinks. We consider how these quantum corrections affect the binding energy of the 2-kink.
 \end{abstract}

\maketitle

\section{Introduction}

Kinks are solitons in one dimension. They are most often used as toy models for solitons in more complicated, higher-dimensional theories. In fact, it was Skyrme who derived the sine-Gordon model as ``a simplified model" of the Skyrme model \cite{skyrme1958non} and thoroughly studied its kink solution, interpreted as particles \cite{skyrme1961particle} (the sine-Gordon model had previously been used in the study of surfaces with constant negative curvature and crystal dislocations). Since then authors have studied more models and their kink solutions, dynamics, and quantum corrections to their masses.

In this paper, we will construct a modfied sine-Gordon theory which supports multi-kinks. Here, we define a multi-kink as a static solution that can be continuously deformed into multiple well-separated 1-kinks, each a solution to the equations of motion. The solutions are non-BPS, so do not satisfy a first-order differential equation. In the following paragraphs, we argue why these non-BPS kinks can serve as better toy models for solitons in higher dimensional theories than BPS kinks.

In higher dimensions, multi-solitons are often the most interesting part of the theory. For example, one has the rich landscape of static multi-skyrmion \cite{Gudnason:2022jkn} and monopole solutions \cite{Sutcliffe:1997ec}. All known one-component kink models satisfy the BPS property \cite{Adam:2016ipc}, and these theories cannot support static stable multi-kinks. Hence a BPS kink theory cannot be used as a toy model for problems involving multi-solitons since they do not support static multi-kinks. The attractive interaction of separated solitons is vital to understanding these multi-solitons, as well as the stability and structure of vortex and skyrmion lattices, which appear in condensed matter theory \cite{abrikosov1957magnetic, yu2010real}. However, the interaction of kinks is repulsive (e.g. sine-Gordon). There can be attractive dynamics, but only between a kink and an anti-kink, whose dynamics will be very different than two kinks.

Another major idea in soliton theory is the collective coordinate approximation, first introduced by Manton for vortex scattering \cite{Manton:1978gf}. Here one constructs a manifold of configurations, sometimes called the configuration space, parametrized by the ``collective coordinates" of the solitons (e.g. their positions). The manifold has a metric and potential induced by the field theory. Soliton dynamics is then described as geodesics on the manifold, crucially depending on the metric and potential. These have been constructed for integrable models such as critically coupled vortices \cite{taubes1980arbitrary, Samols:1991ne}, monopoles \cite{atiyah2014geometry} and instantons \cite{atiyah1994construction}. However, the problem is more complicated for non-BPS theories: there is usually not a canonical choice of collective coordinates and so there are ambiguities in the definition of the configuration space. This ambiguity means there are many different ideas for how to construct CCMs such as the inclusion of shape modes \cite{Manton:2021ipk}, mechanization \cite{Blaschke:2022fxp}, perturbative approaches \cite{Adam:2021gat}, projecting instanton solutions \cite{Adam:2022kla}, pinning \cite{babaev2005semi} and gradient decent curves \cite{Manton:1996ex, Leese:1994hb, Halcrow:2015rvz}. However all the kink studies concern BPS models, or the interaction of kinks with antikinks: not kink-kink dynamics. Constructing collective models for non-BPS multilink scattering may help one understand how to do this systematically for soliton-soliton interactions in higher dimensional models.

There is also significant debate about soliton binding energies. Many papers are motivated to solve the ``binding energy problem" in the Skyrme model: that the classical binding energy of Skyrmions is larger than the binding energy of nuclei by an order of magnitude \cite{Adam:2010fg, Gillard:2015eia, Naya:2018kyi}. However, it has been argued that the classical binding energy is not a good measure of the quantum binding energy. Loop corrections \cite{Meier:1996ng} and the inclusion of ``vibrational modes" \cite{Gudnason:2023jpq} greatly affect skyrmion masses, and have been argued to solve the binding energy problem without using a carefully tuned model, but these ideas require further testing. BPS kink models cannot give much intuition about this problem since their classical binding energies vanish, and there is no multi-kink with a binding energy to measure.

There is a clear need to develop non-BPS kink theories which can serve as toy models for the physically interesting non-BPS models in higher dimensions. Recently one such model has been constructed and studied \cite{Halcrow:2022kfo}, which has also been studied with different motivations \cite{portugues2002intersoliton, Alonso-Izquierdo:2021tqz}. This is a two-component theory in which kinks attract at long-range but repel at short-range. Thus there can be static, stable multi-kinks. However, this model only supports up to two kinks.

In this paper, we introduce a kink theory that supports stable kinks with arbitrary charge $N$, mimicking higher dimensional theories. Our model is a modified version of sine-Gordon theory. We will study the static solutions, create a collective coordinate model for 2-kink dynamics, quantize this and calculate the one-loop quantum corrections. Throughout, we try to answer the questions one would ask about higher-dimensional theories, focusing on the validity of the collective coordinate approximation and binding energies. The broad range of topics that we cover shows the large number of questions one can ask about this new type of kink model.

\section{Stable static multi-kink solutions}

We begin by briefly reviewing \cite{Halcrow:2022kfo}, which details the interaction between kinks. Consider a multi-component scalar theory with Lagrangian
\begin{align}
\label{FFL}
\mathcal{L} &= \tfrac{1}{2} \partial_\mu \Phi_a \partial^\mu \Phi_a - V(\Phi_a) \nonumber\\
&= \tfrac{1}{2} \dot{\Phi}_a \dot{\Phi}_a - \tfrac{1}{2}\partial_x \Phi_a \partial_x \Phi_a - V(\Phi_a) \, .
\end{align}
We will assume that the potential has $N$ minima which we can enumerate. A 1-kink is a static solution of the equations of motion that joins two adjacent minima of the potential. Due to the translational symmetry of $\mathcal{L}$, the kink can be translated. Hence, if we define a base-point configuration with position $X=0$, $\Phi^0(x)$, we can define a kink with position $X$ as $\Phi^X(x) = \Phi^0(x-X)$.

%
%
%
%

Consider a single kink which approaches the vacuum $\Phi^{v_0}$ as $x\to \infty$. Near the vacuum, we can Taylor expand
\begin{align}
\Phi(x)=\Phi^{v_0}+\boldsymbol{\phi}(x) \, .
\end{align}
The tail $\phi_a(x)$ satisfies the Euler-Lagrange equation
\begin{align}
\partial^2_x\phi_a-\partial_a\partial_b V\left(\Phi^{v_0}\right)\phi_b=0
\end{align}
The solution to this equation is given in terms of the eigenvalues $\lambda_n$ and eigenvectors $\boldsymbol{\mu}_n$ of the Hessian $\partial_a\partial_b V(\Phi^{v_0})$,
\begin{align}
\label{tailexp}
\boldsymbol{\phi}=\sum_n a_n \boldsymbol{\mu}_n\,e^{-\sqrt{\lambda_n}x} \, .
\end{align}

Now consider two well-separated kinks with positions ${+X}$ and ${-X}$, denoted by $\Phi^{+X}$ and $\Phi^{-X}$ respectively. Suppose that the two kinks relate three adjacent vacua such that they share one vacuum $\Phi^{v_0}$. A superposition of the kinks is given by
\begin{align} \label{eq:combined}
\Phi(x)=\Phi^{-X}(x)+\Phi^{+X}(x)-\Phi^{v_0}
\end{align}
If $X$ is large, this can be written as
\begin{align}
\Phi(x)&=\Phi^{-X}(x)+\phi^{+X}(x)-\Phi^{v_0}, \quad x << 0\cr
\Phi(x)&=\Phi^{+X}(x)+\phi^{-X}(x)-\Phi^{v_0}, \quad x >> 0
\end{align}
One can then evaluate the static energy of the theory to find the interaction of the two kinks \cite{Halcrow:2022kfo}. It is
\begin{align} \label{eq:interaction_energy}
E^{\text{interaction}}(\Phi)=\left(\partial_x\phi^{-X}_a\phi^{X}_a-\partial_x\phi^{X}_a\phi^{-X}_a\right){\bigg |}_{x=0}
\end{align}
Thus the interaction energy depends only on the tail configurations at the center of mass of the two kinks. This type of argument was first given for planar skyrmion \cite{Schroers:1993yk}, and has since been used for three-dimensional \cite{Feist:2011aa} and magnetic \cite{Barton-Singer:2022rov} skyrmions.

Using \eqref{eq:interaction_energy} one can show that $\phi^4$, $\phi^6$ and sine-Gordon kinks repel, while kinks and antikinks attract. More generally, in one-component scalar field theories, a kink-kink will repel while an antikink-kink will attract and annihilate. As a result, static 2-kink solutions had not been considered until recently. The basic idea in \cite{Halcrow:2022kfo} was to construct solutions which act like a kink-kink in one component and an antikink-kink in another. One component provides repulsion and the other attraction. If these are balanced, a static stable 2-kink can exist.

\section{Modified sine-Gordon theory}

In \cite{Alonso-Izquierdo:2021tqz, Halcrow:2022kfo} a modified $\phi^4$ theory which could support up to two kinks was constructed. However, most higher dimensional soliton theories of interest support an arbitrary number of kinks. To mimic this feature, we will now develop a modified sine-Gordon theory. The new feature of this model compared to other modified sine-Gordon theories \cite{Lund:1976ze, Peyrard:1983rzn, Ferreira:2013xda} is the existence of stable, static multi-kink solutions. Their existence relies crucially on the fact that our model contains more than one field. We will consider the simplest non-trivial case of a two-component theory.

The theory has a Lagrangian of the form \eqref{FFL}. We searched for a modified form of the sine-Gordon potential with an infinite number of discrete vacua and a Hessian that is equal at all vacua. A simple potential that obeys these requirements is 
\begin{align}
  V(\boldsymbol{\Phi})= (1-\cos\Phi_{1})+ \frac{\mu^2}{8}\left(1-\frac{2\Phi_{2}}{m}-\cos\left(\frac{\Phi_{1}}{2}\right)\right)^2\,. 
\end{align}
The potential admits two countably infinite sets of discrete vacua, given by 
\begin{align}
(\Phi_{1},\Phi_{2})&=(4n\pi,0)\,\quad \text{and}\nonumber\\[5pt]
(\Phi_{1},\Phi_{2})&=\left(2(2n+1)\pi,m\right),\,\text{where}\,\,n\in\mathbb Z\,.
\end{align}
Since the minima of our potential, $V(\Phi_1, \Phi_2)$, can be ordered by the $\Phi_1$ coordinate, we can unambiguously define a topological charge of a solution as the difference between its initial and final $\Phi_1$ values. We normalize this value by dividing by $2\pi$. Hence we define an $N$-kink as the minimal energy solution which interpolates between $\Phi_1 = 2\pi n$ and $\Phi_1 = 2\pi(n+N)$. Without loss of generality, we can set $n=0$ and we do so for the rest of the paper. We define the location of the kinks by the preimage of the point halfway between the boundary values.  

First, consider a 1-kink, which interpolates between $(\Phi_1, \Phi_2) = (0,0)$ and $(\Phi_1, \Phi_2) = (2\pi, m)$. The tail of the kink decays exponentially at a rate determined by the Hessian at the vacua \eqref{tailexp}. The Hessian is equal at all vacua, and is given by
\begin{align}
\frac{\partial^2V}{\partial\Phi_a\partial\Phi_b}\Bigm\lvert_{\Phi=(2\pi,m)}=
\begin{pmatrix}
1 & 0\\
0 &  \mu^2/m^2
\end{pmatrix} \, .
\end{align}
Hence the kink decays as
\begin{equation}
    \left(\phi_1, \phi_2\right) \sim \left( a e^{-|x|}, b e ^{-\mu |x|/m} \right)  \, ,
\end{equation}
as $x$ tends to $\pm \infty$.

Now consider the kink-kink interaction where the first kink interpolates between $(0,0)$ and $(2\pi,m)$ and has position $-X$, while the second kink interpolates between $(2\pi,m)$ and $(4\pi,0)$ and has position $X$. The interaction energy again depends on the Hessian, and is given by \eqref{eq:interaction_energy}
\begin{align}
	E^{\text{int}}=a^2e^{-2X}-b^2\frac{\mu}{m}e^{-2\frac{\mu}{m}X} \, ,
\end{align}
The form of the interaction energy reveals that there are competing attractive and repulsive forces, depending on the parameters $m$ and $\mu$. Hence we can expect a stable 2-kink, at least in some parameter range. Since the Hessian is equal at all vacua, the calculation is almost identical if we consider the interaction between a two-kink and a one-kink. The only difference is in the coefficients $a$ and $b$. As such, we expect that if there is a stable 2-kink, there will be a stable 3-kink and so on.

\begin{figure}[H]
    \begin{center}
    \includegraphics[width=\columnwidth]{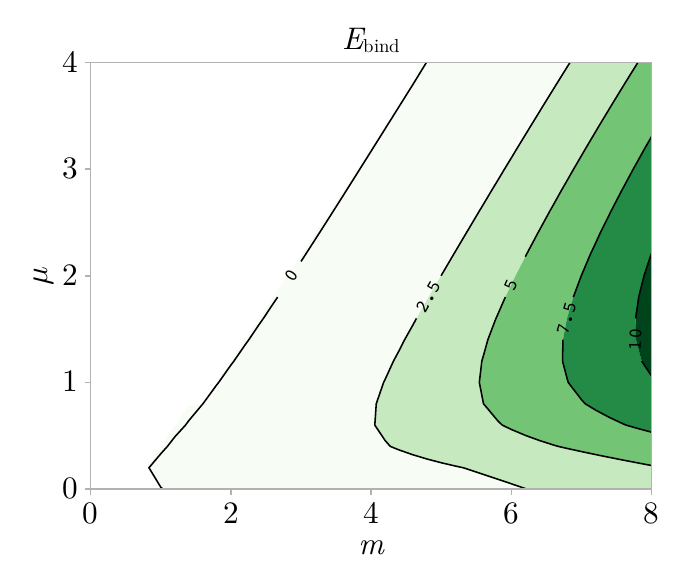}
    \caption{The percentage binding energy per kink for the 2-kink as a function of $m$ and $\mu$.}	\label{fig:2kinkBE}
    \end{center}
\end{figure}

\begin{figure}
	\begin{center}
			\includegraphics[width=\linewidth]{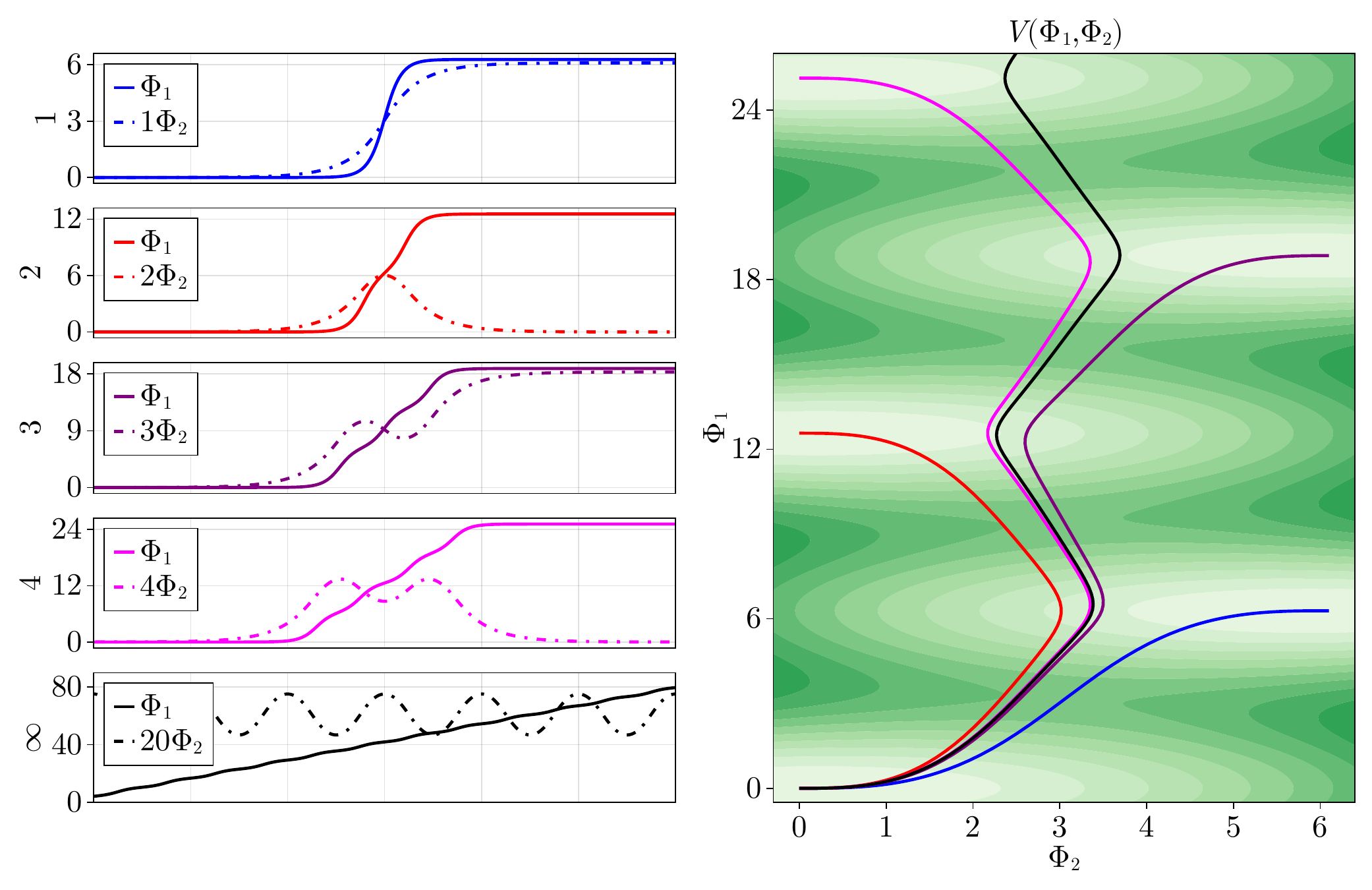}
			\caption{Plots of the 1-, 2-, 3-, 4- and infinite charge kinks, displayed in blue, red, purple, magenta, and black respectively. These are plotted as functions of $x$ (left) and on the target space (right). Light/dark green corresponds to a smaller/larger potential energy. The $N$-kink joins a point with $\Phi_1 = 0$ to a point with $\Phi_1 = 2\pi N$.}	\label{fig:1234inf}
		\end{center}
  \end{figure}

To find $N$-kink solutions, we use numerics. For simplicity, we use a  gradient flow algorithm to find the energy minimising $N$-kink. First, consider the 1- and 2-kinks. We implemented the gradient flow on a grid of 400 points with lattice spacing 0.126 using Python. We considered the following as an initial guess for the 2-kink solution 
\begin{align}
\label{initialconfig}
	\Phi_1&=\pi(2+\tanh(x+X)+\tanh(x-X))\cr
	\Phi_2&=me^{-x^2}
\end{align}
With the parameters $\mu=2, m=6.1$, the energy of the 1-kink was found to be $E_1=10.24$, whereas the 2-kink has energy $E_2=19.44$. Hence the 2-kink is energetically favoured over two 1-kinks. The percentage binding energy per soliton is
\begin{align}
	E_{\text{bind}}=\frac{2E_1-E_2}{2E_1}\times 100\%=5.1\% \, ,
\end{align}
The binding energy depends on the parameters $m$ and $\mu$. We find this dependence and plot it in Figure \ref{fig:2kinkBE}. The plot can be used to choose the parameters to generate a model with the desired binding energy. Later we will study a variety of models, including the model with $\mu=2, m=4$. This is much closer to BPS with percentage binding energy per soliton of only $0.8\%$.
There are also stable kinks for larger $N$. The 1-, 2-, 3-, 4- and infinite charge kinks are displayed in Figure \ref{fig:1234inf}. We plot the fields $\Phi_a(x)$ (left) and also plot these on the potential landscape (right). Note that the finite-$N$ kinks terminate at a minimum, while the infinite kink weaves past all these points. The infinite kink was calculated by putting the model on a circle of length $L$ and using shifted periodic boundary conditions. These are simple for a 2-kink:
\begin{equation}
    \Phi_1(x + L) = 4\pi + \Phi_1(x), \quad \Phi_2(x+L) = \Phi_2(x) \, .
\end{equation}
We vary over  $L$ to find the minimum energy chain. The numerically obtained solution was then repeated on a larger grid, and this is shown in Figure \ref{fig:1234inf} (bottom left).

\onecolumngrid

\twocolumngrid
  
\begin{figure}[H]
    \begin{center}
            \includegraphics[width=\columnwidth]{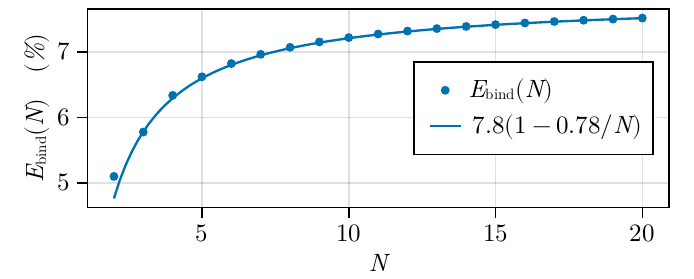}
            \caption{The percentage binding energy per kink, for $N=1-20$, for $\mu=2$ and $m=6.1$.}	
            \label{fig:bind}
    \end{center}
\end{figure}

The classical percentage binding energy per kink of an $N$-kink is given by the formula
\begin{equation}
    E_\text{bind}(N) = \frac{NE_1 - E_N}{NE_1}  \, .
\end{equation}
This is an important quantity in the Skyrme model, where the formula gives the classical binding energy of nuclei per nucleon. We calculate this value for $N=1-20$ in our model, and plot it in Figure \ref{fig:bind}. Similar to nuclei, kink binding energies asymptote to a constant value, in this case $7.8\%$. We fit points to a polynomial decay and find that the classical binding energy per kink goes as $1/N$.

\section{Collective coordinate model}

\subsection{Making the model}
In this section we construct a one-dimensional collective coordinate model (CCM) for two kinks, in which we use the kink positions $\pm X$ as the modulus. The CCM is described by a Lagrangian of the form
\begin{align}
\label{CCML}
L=\frac{1}{2}g_{XX}(X)\dot X^2-U(X)
\end{align}
where the metric $g_{XX}$ on the one-dimensional moduli space is 
\begin{align}
\label{metricformula}
    g_{XX}(X)=\int_{-\infty}^{\infty}\left[\left(\frac{\partial\Phi_1}{\partial X}\right)^2+\left(\frac{\partial\Phi_2}{\partial X}\right)^2\right]dx
\end{align}
and the potential $U(X)$ is   
\begin{align}
\label{potentialformula}
U(X)=\int_{-\infty}^{\infty}\left(\frac{1}{2}\left(\frac{\partial\Phi_1}{\partial x}\right)^2+\frac{1}{2}\left(\frac{\partial\Phi_2}{\partial x}\right)^2+V(\Phi)\right)dx
\end{align}
Let us again consider two one-kinks, one that interpolates from $(0,0)$ to $(2\pi,m)$ and another that interpolates from $(2\pi,m)$ to $(4\pi,0)$. We define the position of these kinks $X$ and $-X$ as the preimage where the first component takes value at the midpoint, i.e. we pin the position of the kinks to $X$ and $-X$ such that 
\begin{align}
\label{pinningeqn}
\Phi_1(-X)=\pi,\quad \Phi_1(X)=3\pi \, .
\end{align}
For a given separation $2X$ between the two kinks, we obtained the configuration $(\Phi_1,\Phi_2)$ that minimizes the energy \eqref{potentialformula} using gradient flow.  We again implemented the flow on a grid with 400 points and lattice spacing 0.126 and used the configuration in \eqref{initialconfig} as our initial guess. We did this for $0.5<X<13$ with $X$ fixed at each $x$ lattice point, generating around 100 configurations for the model. This generates the configuration space, parametrized by $\Phi(x,X)$. We used the central difference method to calculate the derivatives of $\Phi$ with respect to both the kink position $X$ and the spatial coordinate $x$ in \eqref{metricformula} and \eqref{potentialformula}.

\begin{figure}[H]
 \begin{center}
		\includegraphics[width=\columnwidth]{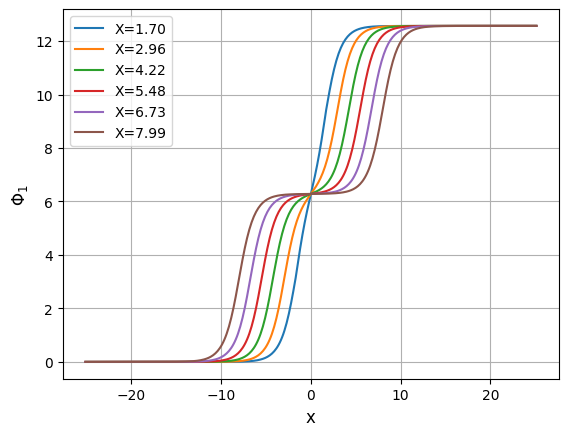}
	
		\includegraphics[width=\columnwidth]{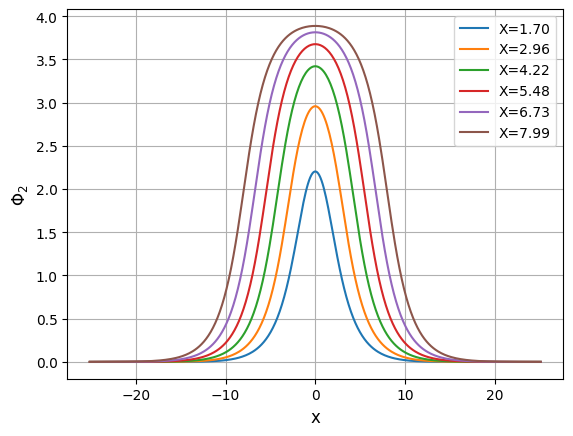}
		\caption{Energy minimising $\Phi_1$ (top) and $\Phi_2$ (bottom) for different values of $X$, for $\mu=2$ and $m=4$.}	\label{stable phi2MSG}
  \end{center}
  \end{figure}

\begin{figure}[H]
	\begin{center}
		
			\includegraphics[width=\columnwidth]{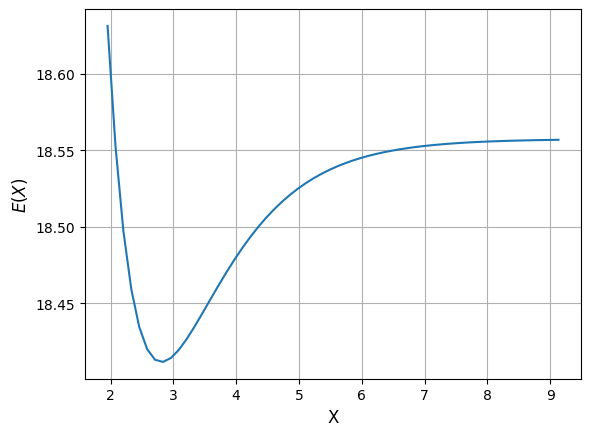}
			\caption{Energy as a function of $X$ for $\mu=2$ and $m=4$.}	\label{Energy MSG}
		\end{center}
  \end{figure}

\begin{figure}[H]
	\begin{center}
		
			\includegraphics[width=0.9\columnwidth]{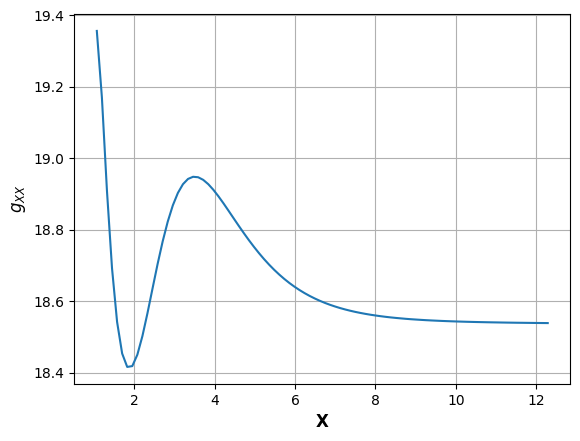}
			\caption{Metric $g_{XX}$ as a function of $X$, for $\mu=2$ and $m=4$.}	\label{metric MSG}
		\end{center}
  \end{figure}

Several configurations for different $X$ are plotted in Fig.  \ref{stable phi2MSG}. The resulting energy of these configurations as a function of the separation is plotted in Fig. \ref{Energy MSG}. We see that the energy as a function of the separation has a minimum at a finite non-zero $X$, agreeing with the field theory expectation that the two-kink is energetically favored over two one-kinks. The energy increases sharply when the kinks are brought closer together, but only increases slowly as they are moved further apart. The metric is plotted in Fig. \ref{metric MSG}, and has two critical points.


\subsection{Kink dynamics}

Our collective coordinate model can be used to simulate kink-kink dynamics. We will first study the full field time evolution of the kinks by solving the equations of motion in the field theory. We will then go on to see how well these are modelled by our CCM.

The full field dynamics is given by the Euler-Lagrange equation of motion that follow from \eqref{FFL}
\begin{equation}
\ddot{\Phi}_a = \frac{\partial^2 \Phi_a}{\partial x^2} - \frac{\partial V}{\partial \Phi_a} \, .
\end{equation}
To solve this, we introduce the field velocity $\Psi$, 
\begin{align}
\dot{\Phi}_a&=\Psi_a\\
    \dot{\Psi}_a&=\frac{\partial^2 \Phi_a}{\partial x^2} - \frac{\partial V}{\partial \Phi_a}
    \end{align}
We solve these numerically using the Leapfrog method. We update $\Phi$ and $\Psi$ as follows
\begin{align}
    \Phi_{n+1}=\Phi_{n}+\Psi_{n}dt+\frac{1}{2}b_{n}dt^2\\
    \Psi_{n+1}=\Psi_{n}+\frac{1}{2}\left(b_{n}+b_{n+1}\right)dt
\end{align}
where
\begin{align}
    b_{n}=  \frac{\partial^2 \Phi_{n}}{\partial x^2} - \frac{\partial V_{n}}{\partial \Phi_{n}} \, .
\end{align}
From the time evolution of $\Phi$, we obtain the time evolution of $X$ using \eqref{pinningeqn}. The initial velocity field is generated by
\begin{equation}
    \Psi_0(x) = \frac{\partial}{\partial t} \Phi(X-vt,x)\rvert_{t=0} = -v\frac{\partial \Phi(X,x) }{\partial X} \, .
\end{equation}

In the collective coordinate approximation, dynamics is dictated by the equation of motion that follows from \eqref{CCML}
\begin{equation}
    \ddot{X}+ \frac{1}{2g}\left( \left(\partial_X g(X)\right) \dot{X}^2 + 2\partial_X U(X) \right) = 0 
\end{equation}
To solve this, we introduce $v = \dot{X}$. The equations become
\begin{align}
\dot{X} &= v \\
\dot{v} &= -\frac{1}{2g}\left((\partial_X g) v^2 +2\partial_X U \right)
\end{align}
Starting with the initial data $(X, v) = (X_0, v_0)$, we evolve these equations numerically using the Leapfrog method, updating $X$ and $v$ as follows
\begin{align}
    X_{n+1}=X_{n}+v_{n}dt+\frac{1}{2}a_{n}dt^2\\
    v_{n+1}=v_{n}+\frac{1}{2}\left(a_{n}+a_{n+1}\right)dt
\end{align}
where
\begin{align}
    a_{n}= -\frac{1}{2g}\left( \left(\partial_X g(X_n)\right) v_{n}^2+2\partial_X U(X_n) \right)
\end{align}
Here the derivatives of the metric $g_{XX}$ and the potential $U$ with respect to the kink position $X$ are calculated using a cubic spline.

\begin{figure}[H]
 \begin{center}
		\includegraphics[width=\columnwidth]{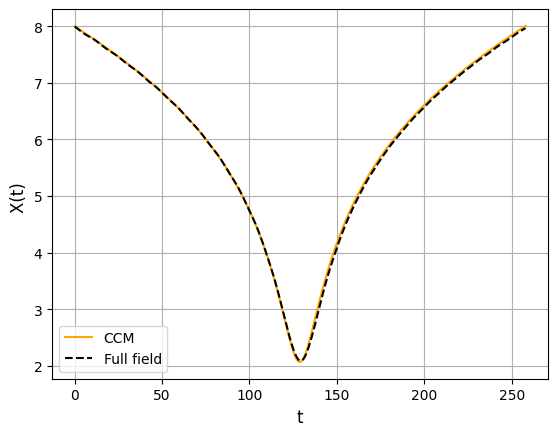}
		\caption{$X(t)$ for initial velocity $v=0.02$ and initial position 8 for $\mu=2$ and $m=4$.}	\label{v=0.02}
	\end{center}
  \end{figure}

\begin{figure}[H]
 \begin{center}
		\includegraphics[width=\columnwidth]{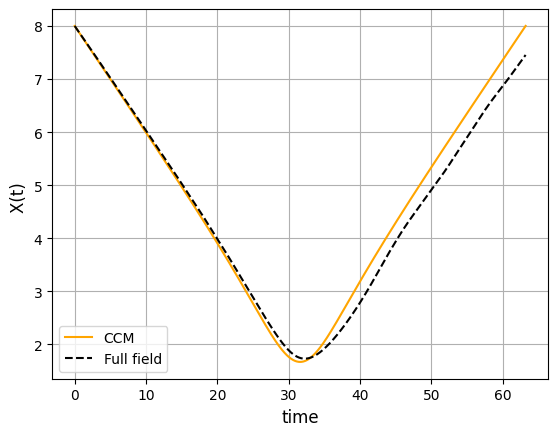}
		\caption{$X(t)$ for initial velocity $v=0.2$ and initial position 8 for $\mu=2$ and $m=4$.}	\label{v=0.2}
	\end{center}
  \end{figure}

We now generate $X(t)$, obtained from the CCM as well as from the full field theory for different values of initial velocity. We first consider trajectories where $X(0) = 8$ and $v = 0.02, 0.2$. We interrupt the dynamics when $X(t)>8$ in the CCM. The results are plotted in Figs. \ref{v=0.02} and \ref{v=0.2} for the model with parameters $\mu=2$ and $m=4$. This model is close to the BPS model, and we obtain a very good approximation for small velocities, with the approximation becoming worse as we increase the velocity. This is expected. As the configuration passes through the minimum of the potential, there is a large energy transfer from potential to kinetic energy and back again. In the full field theory, the 2-kink can radiate, losing energy. Hence the outgoing kinks have a phase shift and are slower than the incoming kinks. In the collective coordinate model, the kink cannot radiate and the outgoing kinks have the same velocity as the incoming ones. 

\begin{figure}[H]
 \begin{center}
		\includegraphics[width=\columnwidth]{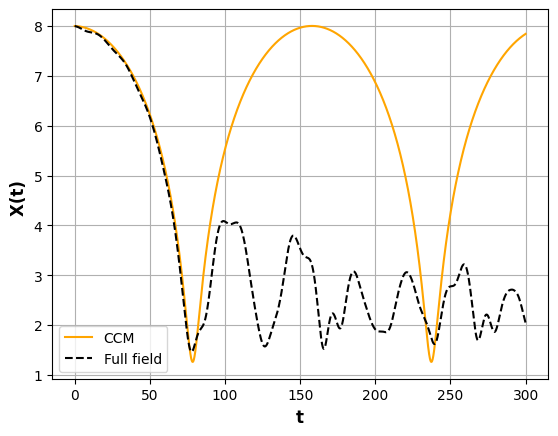}
		\caption{$X(t)$ for initial velocity $v=0.01$ and initial position 8 for $\mu=2$ and $m=6.1$.}	\label{v=0.001}
	\end{center}
  \end{figure}

The approximation appears to break down when we move away from near-BPS models. In Fig. \ref{v=0.001} we consider the parameters $\mu=2$ and $m=6.1$. We display this for a longer time than the other plots, to better understand the behavior of the field theory. At large $t$, the approximation becomes a lot less accurate. Note that the ``meson" mass in the second field is $\mu/m$ so that as $m$ increases, it costs less and less energy to excite radiation. Hence radiation becomes more important as $m$ increases. Here, the two kinks radiate their kinetic energy away and cannot escape from the potential well. Again, since the CCM cannot radiate, it cannot model this behavior. Hence the CCM does not capture this main qualatative feature of the dynamics. 

However, note that the model does capture the correct behavior until the 2-kink approaches the bottom of the potential. Thus the model does seem to capture the potential and curvature of the configuration space. Hence it might not be our model that fails, but the geodesic approximation itself. The approximation should fail when velocities and radiation are high, so it makes sense that it fails when the kinks are at the bottom of the potential well, where their velocity is maximal. Another possibility is that our CCM model is too simple. Other authors have considered CCMs with multiple collective coordinates, one which usually describes an internal shape mode of the kinks \cite{ Manton:2021ipk, Adam:2021gat}. We could extend our model by including this shape mode for one or each of the kinks, and see if the approximation becomes significantly better.

\begin{figure}[H]
 \begin{center}
		\includegraphics[width=\columnwidth]{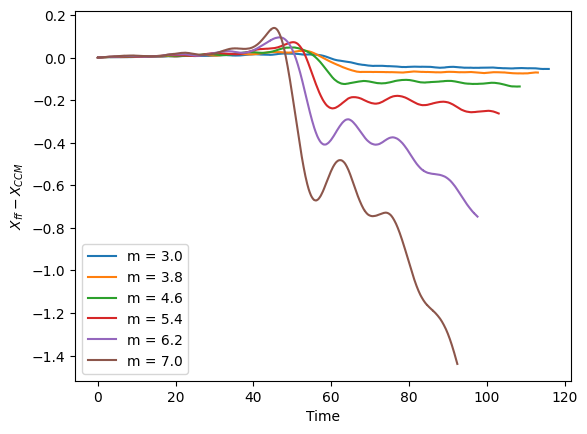}
		\caption{The difference between the full field calculation $X_{ff}(t)$ and the collective coordinate model $X_{CCM}(t)$ in the case $X(0) = 8, v=0.1$ for $\mu=2$ and a variety of $m$.}	\label{fig:errors}
	\end{center}
  \end{figure}

The geodesic approximation relies on two assumptions: that the velocity (and hence energy transfer to radiation) is small and that the theory is close to BPS. We now probe the second of these assumptions in more detail. Consider the scattering of two kinks with initial positions $X=8$ and $X=-8$ with velocity $v=0.1$, the model parameters $\mu=2$ fixed and $m$ varying. We construct the CCM, simulate the CCM dynamics and the full field dynamics for a variety of $m$. As $m$ increases we move further from a BPS theory. The difference between CCM and full field dynamics for the various models are plotted in Fig. \ref{fig:errors}. We see that for $m<5.4$ the CCM captures the main features of the model: although a large error is generated as the model passes near the bottom of the potential, the outgoing kinks have approximately constant velocity and the error remains approximately constant after the bounce. However, when $m>5.4$ the qualitative behavior changes: this is when the 2-kink gets trapped in the potential well, as seen in Fig. \ref{v=0.001}.

Overall, we have successfully constructed collective coordinate models for non-BPS 2-kink dynamics. Using these we have generated trajectories and compared the CCM dynamics with full field dynamics. As the initial velocity increases, and as the model becomes further from BPS, the approximation gets worse, as expected. We believe the biggest problem is that the CCM model cannot radiate. Radiation is also important for a variety of other kink models, and authors have begun to incorporate radiation \cite{Navarro-Obregon:2023hqe} to CCMs, although it is complicated. Our results show that radiation becomes more important as we move further from BPS theories. The results suggest that accurate CCMs of, e.g., the Skyrme model may be more difficult to construct than previously thought.
  
\subsection{Quantum kinks}

The classical Lagrangian of our collective coordinate model of two kinks takes the form
\begin{equation}
    \mathcal{L} = \tfrac{1}{2}g(X)\dot{X}^2 - U(X) \, .
\end{equation}
We can apply canonical quantisation to find the semiclassical Hamiltonian
\begin{equation}
    \hat{\mathcal{H}} = -\frac{\hbar^2}{2}\Delta + U(X) \, .
\end{equation}
Naively, the kinetic operator is just $g^{-1}\partial_X^2$, but the metric makes things more complicated: $\Delta$ is actually the Laplace-Beltrami operator. In one dimension it takes the form
\begin{equation}
    \Delta  = \frac{1}{\sqrt{g}}\partial_X\left( \frac{1}{\sqrt{g}}\partial_X\right)  =\frac{1}{g}\left( \partial_X^2 - \frac{\partial_X g}{2g}\partial_X \right) \, .
\end{equation}

In the semiclassical approximation, bound states solve the Schr\"odinger equation
\begin{equation}
    -\frac{\hbar^2}{2}\Delta \Psi + U(X) \Psi = E \Psi \, .
\end{equation}
Before solving the equation exactly, we can make an approximate solution using the harmonic approximation. Here, we ignore the first derivative term in the Laplace-Beltrami operator and approximate the potential by a quadratic. The Schr\"odinger equation becomes
\begin{equation} \label{eq:harmSchro}
    -\frac{\hbar^2}{2g}\partial_X^2 \Psi + \frac{1}{2}
    \omega^2 X^2 \Psi = E \Psi \, ,
\end{equation}
where $\omega$ is the first normal mode frequency, which can be found using the normal mode equations (discussed in the next section). Equation \eqref{eq:harmSchro} has ground state solution with $E = \tfrac{1}{2}\hbar \omega$. In our case, for $\mu=2, m=6.1$, the frequency is equal to $\omega = 0.2072$.  In units with $\hbar=1$, the ground state has energy $0.1036$. The quantum 1-kink has no bound mode, and so in a harmonic approximation its energy is simply the classical energy. Hence we can calculate the quantum binding energy in the harmonic approximation. We find that
\begin{align}
E^\text{Harmonic}_1 = 10.2422, \quad E_2^\text{Harmonic} =19.5440 \, ,\\
\implies E_\text{bind}^\text{Harmonic} = 4.59\% \, .
\end{align}
The inclusion of quantum energy increases the energy of the 2-kink and so decreases the binding energy.

\begin{figure}
    \centering
    \includegraphics[width=\columnwidth]{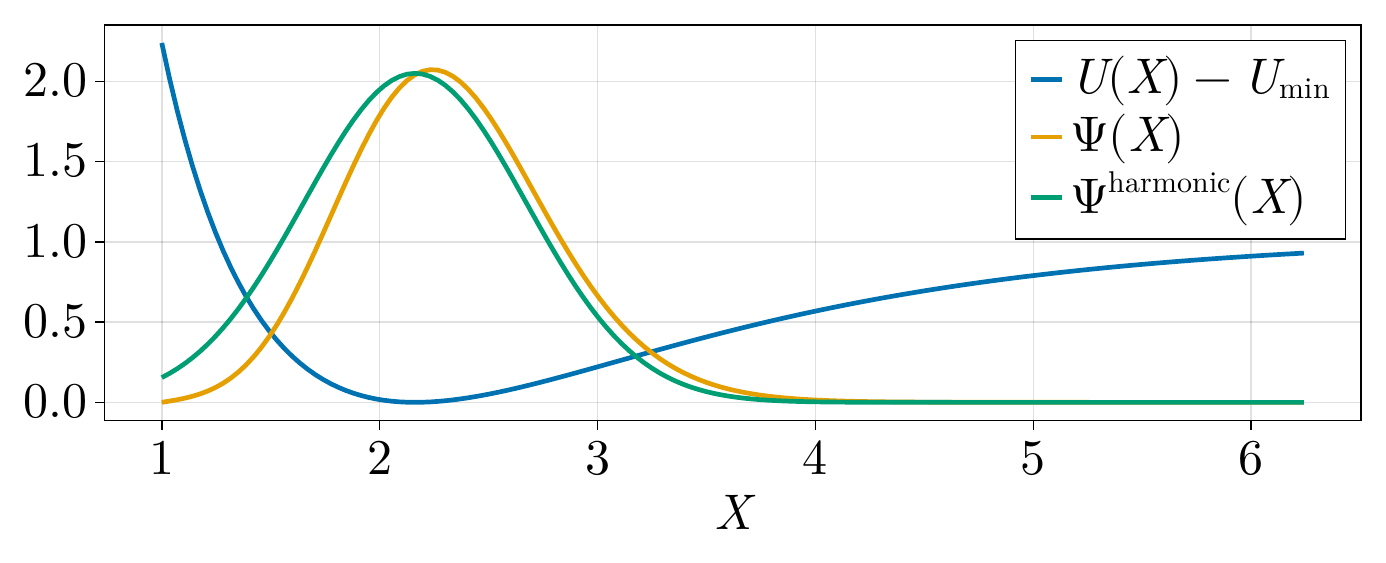}
    \caption{The ground state wavefunction of the 2-kink in the collective coordinate approximation when $m=6.1, \mu=2$, in a harmonic approximation (green) and the true solution (orange). We also plot the shifted potential energy (blue).}
    \label{fig:wavefunction}
\end{figure}

To solve the Schr\"odinger equation beyond the harmonic approximation, giving the true wavefunction, we take an initial random guess $\Psi_0(X)$ and evolve it using
\begin{equation} \label{eq:SchroEvolve}
    \dot{\Psi} = -\hat{\mathcal{H}}\Psi\, , \quad \Psi^2 = 1 \, .
\end{equation}
The ground state is the late-time solution of \eqref{eq:SchroEvolve}. In one-dimension there is always a ground state solution. We calculate it for $\hbar=1$, and it is shown in Figure \ref{fig:wavefunction}, alongside the harmonic solution. Both are concentrated near the minimum of $U$. The main difference between wavefunctions is that the boundary at the left squeezes the true wavefunction, moving it slightly rightwards. Due to this squeeze, the energy of the anharmonic wavefunction is $0.1107$, larger than the harmonic approximation. This affects the binding energy, which is now equal to
\begin{align}
\implies E_\text{bind}^\text{Quantum anharmonic} = 4.55\% \, .
\end{align}
Hence the anharmonticity has decreased the binding energy of the quantum 2-kink by a small amount.

We repeat the calculation for the near-BPS case, $\mu=2, m=4$. Here the quantum energy is $0.059$ in the harmonic approximation and $0.053$ in the anharmonic case. The anharmonic quantum energy is less. The percentage binding energy per kink goes from $0.79\%$ to $0.47\%$ for the harmonic approximation (or $0.50\%$ in the anharmonic case). So even though the quantum correction is smaller, the effect on the percentage binding energy is much larger.

\section{One-loop correction}

We can calculate the one-loop correction to kink masses using the formula of Cahill, Comtet and Glauber \cite{Cahill:1976im}. This can be written in terms of fluctuations $\epsilon_n(x)$ around the classical kink solutions $\Phi^{(N)}$, where $N$ represents the kink number. To find their equations of motion, first write the field as
\begin{align}
 \Phi(x,t) = \Phi^{(N)}(x) + \sum_n \epsilon_n(x) e^{i \omega t} \, ,
\end{align}
where $\omega$ are the normal mode frequencies. These satisfy the normal mode equation
\begin{equation} \label{eq:flucs}
\left(-\delta_{ab} \partial_x^2 + V_{ab}[\Phi^{(N)}]\right) \epsilon_b = \omega^2 \epsilon_a \, ,
\end{equation}
where $V_{ab}$ is given by the Hessian
\begin{equation}
\begin{pmatrix} \frac{ \mu^2(m-2\Phi_2) \cos(\Phi_1/2) - m(\mu^2-16)\cos(\Phi_1)}{16m} & -\frac{\mu^2 \sin(\Phi_1/2)}{4m} \\
    -\frac{\mu^2 \sin(\Phi_1/2)}{4m} & \frac{\mu^2}{m^2} \end{pmatrix} \, ,
\end{equation}
which depends on the solution $\Phi$. The fluctuations form a complete orthogonal basis of $\mathbb{R}$.

The $N$-kink mass is only finite when compared to the mass of the vacuum, $N=0$. Denote the vacuum fluctuations and frequencies with a tilde, so that
\begin{equation}
\left(-\delta_{ab} \partial_x^2 + V_{ab}[\Phi^{(0)}]\right) \tilde{\epsilon}_b = \tilde{\omega}^2 \tilde{\epsilon}_a \, .
\end{equation}
The one loop mass correction of the $N$-kink is then
 \begin{align} \label{eq:oneloop}
\delta m = -\frac{1}{4}\sum_{n,m} \braket{\epsilon_n|\tilde{\epsilon}_m}^2  \left(\frac{\omega_n^2 }{\tilde{\omega}_m} -2 \omega_n + \tilde{\omega}_n \right) , 
\end{align}
where
\begin{equation}
\braket{ \boldsymbol{\epsilon}_n | \boldsymbol{\epsilon}_m } = \int \boldsymbol{\epsilon}_n(x)\cdot \boldsymbol{\epsilon}_m \, dx \, .
\end{equation}

Using the formula \eqref{eq:oneloop}, we can calculate the one-loop mass correction of the kink by solving the eigenvector problem \eqref{eq:flucs}. We do this for 1-kink and the 2-kink for $\mu=2, m=6.1$, on a 600 point grid with a 1200 dimensional eigenbasis $\{ \boldsymbol{\epsilon}_n \}$. We find that the mass correction for the 1-kink is $-0.297$. The mass correction is similar to the analytically known value when $\mu=0$, which is $-1/\pi = -0.318$.

The mass correction for the 2-kink is $-0.486$, which is less than twice the mass correction of the 1-kink. Hence the percentage binding energy of the 2-kink, taking the one-loop correction into account, is $4.71\%$. Hence the loop correction binding energy is very similar to the harmonic binding energy, only slightly smaller: $4.71\%$ compared to $4.59\%$.

We repeat the calculation when $\mu =2, m=4$. Here, the loop corrections for the one- and two- kinks are $-0.302$ and $-0.547$ respectively. Again, the two-kink correction is smaller than two times the one-loop correction. So again, the binding energy decreases due to the loop correction. The percentage binding energy per kink becomes $0.499\%$.

\section{Conclusion and Further Work}

In this paper we have constructed a modified sine-Gordon model which supports stable static multi-kink solutions. Unlike usual sine-Gordon, the model is not integrable or BPS and the solutions have a binding energy. This makes the theory a good toy model for physical systems such as those found in condensed matter and nuclear theory. We believe our model is an excellent testing place for ideas in classical and quantum soliton dynamics. As such, we studied a variety of problems in the new model.

First, we calculated kink solutions for $N=1-20$ and found that the binding energy per kink asymptotes to a constant value, similar to what is seen in nuclear matter. We then constructed a collective coordinate model of 2 kinks using pinning. This is the first time a potential and metric have been generated on a soliton configuration space using this method. In some sense, the method is more general than that used in recent works \cite{Manton:2021ipk}, as it does not rely on features of the 1-soliton to generate approximate 2-soliton fields. We used the model to test the collective coordinate approximation for soliton dynamics against full field theory finding that, as expected, the model works better when the model is closer to BPS. We also used the collective coordinate model to generate a quantum 2-kink solution in semi-classical harmonic and anharmonic approximations. Finally, we calculated the one-loop correction to the 1- and 2-soliton masses. 

Overall, we calculated the binding energy four times: classically, harmonically and anharmonically in a collective coordinate approximation, and using one-loop corrections. The results of these calculations are shown in Table \ref{tab:binding_results}. The fact that the results are different highlights that quantum binding energy is highly dependent on the method used to calculate it. In the Skyrme model, authors have calculated the binding energy classically \cite{Adam:2010fg}, in a harmonic \cite{Gudnason:2023jpq, Walet:1996he} and anharmonic \cite{Leese:1994hb} collective coordinate approach and including loop corrections (for 2D skyrmions) \cite{Walliser:1999ug}. The calculations in higher dimensions are very difficult, and so our model is an excellent tool for probing questions about quantum soliton masses in a simple setting.

\begin{table}[h!]
    \centering
    \begin{tabular}{l|ccc| ccc}
        & $ m = 6.1$ & & & $m=4$ & & \\
         Method&  $E_1$&  $E_2$& $E_\text{bind} $&  $E_1$&  $E_2$& $E_\text{bind} $\\ \hline
         Classical&  $10.242$&  $19.440$& $5.10$& 9.284 & 18.420 & 0.79 \\
         Harmonic&  $10.242$&  $19.544$& $4.59$& 9.284 & 18.479 & 0.47 \\
         Anharmonic&  $10.242$&  $19.552$& $4.55$& 9.284 & 18.474 & 0.50 \\
         One-loop&  $9.946$&  $18.954$& $4.71$& 8.982 & 17.874 & 0.50 \\
    \end{tabular}
    \caption{Energies and binding energies using different approximations, for the 2-kink with $\mu=2, m=6.1$ and $m=4$.}
    \label{tab:binding_results}
\end{table}

There are many obvious generalisations of the work done here: one could generate collective coordinates for more than 2 kinks, calculate higher loop corrections \cite{Evslin:2021vgk} or ask how the binding energy calculation changes when considering more kinks. It would be interesting to add ``isospin" coordinates to the model, similar to the complex sine-Gordon model \cite{Lund:1976ze}, where solitons have additional internal symmetries. The quantisation of this type of model would be closer to higher-dimensional models, which often have internal degrees of freedom. We have modified the normal sine-Gordon theory so that there are bound multi-kinks. But there are various other models that one could modify and study, depending on their interest.
\\ \\
\noindent \textbf{Acknowledgements:} We thank Andrzej Wereszczyński for comments on the manuscript. C.H. thanks Kathy Hubbard for suggesting the use of a cubic spline. R.R.J and A.N thank the Kerala Theoretical Physics Initiative for facilitating their collaboration. C.H. is supported by the Carl Trygger Foundation through the grant CTS 20:25. 

\bibliography{main.bib}

\begin{thebibliography}{39}%
\makeatletter
\providecommand \@ifxundefined [1]{%
 \@ifx{#1\undefined}
}%
\providecommand \@ifnum [1]{%
 \ifnum #1\expandafter \@firstoftwo
 \else \expandafter \@secondoftwo
 \fi
}%
\providecommand \@ifx [1]{%
 \ifx #1\expandafter \@firstoftwo
 \else \expandafter \@secondoftwo
 \fi
}%
\providecommand \natexlab [1]{#1}%
\providecommand \enquote  [1]{``#1''}%
\providecommand \bibnamefont  [1]{#1}%
\providecommand \bibfnamefont [1]{#1}%
\providecommand \citenamefont [1]{#1}%
\providecommand \href@noop [0]{\@secondoftwo}%
\providecommand \href [0]{\begingroup \@sanitize@url \@href}%
\providecommand \@href[1]{\@@startlink{#1}\@@href}%
\providecommand \@@href[1]{\endgroup#1\@@endlink}%
\providecommand \@sanitize@url [0]{\catcode `\\12\catcode `\$12\catcode
  `\&12\catcode `\#12\catcode `\^12\catcode `\_12\catcode `\%12\relax}%
\providecommand \@@startlink[1]{}%
\providecommand \@@endlink[0]{}%
\providecommand \url  [0]{\begingroup\@sanitize@url \@url }%
\providecommand \@url [1]{\endgroup\@href {#1}{\urlprefix }}%
\providecommand \urlprefix  [0]{URL }%
\providecommand \Eprint [0]{\href }%
\providecommand \doibase [0]{http://dx.doi.org/}%
\providecommand \selectlanguage [0]{\@gobble}%
\providecommand \bibinfo  [0]{\@secondoftwo}%
\providecommand \bibfield  [0]{\@secondoftwo}%
\providecommand \translation [1]{[#1]}%
\providecommand \BibitemOpen [0]{}%
\providecommand \bibitemStop [0]{}%
\providecommand \bibitemNoStop [0]{.\EOS\space}%
\providecommand \EOS [0]{\spacefactor3000\relax}%
\providecommand \BibitemShut  [1]{\csname bibitem#1\endcsname}%
\let\auto@bib@innerbib\@empty
\bibitem [{\citenamefont {Skyrme}(1958)}]{skyrme1958non}%
  \BibitemOpen
  \bibfield  {author} {\bibinfo {author} {\bibfnamefont {T.}~\bibnamefont
  {Skyrme}},\ }\href {\doibase https://doi.org/10.1098/rspa.1958.0183}
  {\bibfield  {journal} {\bibinfo  {journal} {Proceedings of the Royal Society
  of London. Series A. Mathematical and Physical Sciences}\ }\textbf {\bibinfo
  {volume} {247}},\ \bibinfo {pages} {260} (\bibinfo {year}
  {1958})}\BibitemShut {NoStop}%
\bibitem [{\citenamefont {Skyrme}(1961)}]{skyrme1961particle}%
  \BibitemOpen
  \bibfield  {author} {\bibinfo {author} {\bibfnamefont {T.}~\bibnamefont
  {Skyrme}},\ }\href {\doibase https://doi.org/10.1098/rspa.1961.0115}
  {\bibfield  {journal} {\bibinfo  {journal} {Proceedings of the Royal Society
  of London. Series A. Mathematical and Physical Sciences}\ }\textbf {\bibinfo
  {volume} {262}},\ \bibinfo {pages} {237} (\bibinfo {year}
  {1961})}\BibitemShut {NoStop}%
\bibitem [{\citenamefont {Gudnason}\ and\ \citenamefont
  {Halcrow}(2022)}]{Gudnason:2022jkn}%
  \BibitemOpen
  \bibfield  {author} {\bibinfo {author} {\bibfnamefont {S.~B.}\ \bibnamefont
  {Gudnason}}\ and\ \bibinfo {author} {\bibfnamefont {C.}~\bibnamefont
  {Halcrow}},\ }\href {\doibase 10.1007/JHEP08(2022)117} {\bibfield  {journal}
  {\bibinfo  {journal} {JHEP}\ }\textbf {\bibinfo {volume} {08}},\ \bibinfo
  {pages} {117} (\bibinfo {year} {2022})},\ \Eprint
  {http://arxiv.org/abs/2202.01792} {arXiv:2202.01792 [hep-th]} \BibitemShut
  {NoStop}%
\bibitem [{\citenamefont {Sutcliffe}(1997)}]{Sutcliffe:1997ec}%
  \BibitemOpen
  \bibfield  {author} {\bibinfo {author} {\bibfnamefont {P.~M.}\ \bibnamefont
  {Sutcliffe}},\ }\href {\doibase 10.1142/S0217751X97002504} {\bibfield
  {journal} {\bibinfo  {journal} {Int. J. Mod. Phys. A}\ }\textbf {\bibinfo
  {volume} {12}},\ \bibinfo {pages} {4663} (\bibinfo {year} {1997})},\ \Eprint
  {http://arxiv.org/abs/hep-th/9707009} {arXiv:hep-th/9707009} \BibitemShut
  {NoStop}%
\bibitem [{\citenamefont {Adam}\ and\ \citenamefont
  {Santamaria}(2016)}]{Adam:2016ipc}%
  \BibitemOpen
  \bibfield  {author} {\bibinfo {author} {\bibfnamefont {C.}~\bibnamefont
  {Adam}}\ and\ \bibinfo {author} {\bibfnamefont {F.}~\bibnamefont
  {Santamaria}},\ }\href {\doibase 10.1007/JHEP12(2016)047} {\bibfield
  {journal} {\bibinfo  {journal} {JHEP}\ }\textbf {\bibinfo {volume} {12}},\
  \bibinfo {pages} {047} (\bibinfo {year} {2016})},\ \Eprint
  {http://arxiv.org/abs/1609.02154} {arXiv:1609.02154 [hep-th]} \BibitemShut
  {NoStop}%
\bibitem [{\citenamefont {Abrikosov}(1957)}]{abrikosov1957magnetic}%
  \BibitemOpen
  \bibfield  {author} {\bibinfo {author} {\bibfnamefont {A.~A.}\ \bibnamefont
  {Abrikosov}},\ }\href@noop {} {\bibfield  {journal} {\bibinfo  {journal}
  {Journal of Physics and Chemistry of Solids}\ }\textbf {\bibinfo {volume}
  {2}},\ \bibinfo {pages} {199} (\bibinfo {year} {1957})}\BibitemShut {NoStop}%
\bibitem [{\citenamefont {Yu}\ \emph {et~al.}(2010)\citenamefont {Yu},
  \citenamefont {Onose}, \citenamefont {Kanazawa}, \citenamefont {Park},
  \citenamefont {Han}, \citenamefont {Matsui}, \citenamefont {Nagaosa},\ and\
  \citenamefont {Tokura}}]{yu2010real}%
  \BibitemOpen
  \bibfield  {author} {\bibinfo {author} {\bibfnamefont {X.}~\bibnamefont
  {Yu}}, \bibinfo {author} {\bibfnamefont {Y.}~\bibnamefont {Onose}}, \bibinfo
  {author} {\bibfnamefont {N.}~\bibnamefont {Kanazawa}}, \bibinfo {author}
  {\bibfnamefont {J.~H.}\ \bibnamefont {Park}}, \bibinfo {author}
  {\bibfnamefont {J.}~\bibnamefont {Han}}, \bibinfo {author} {\bibfnamefont
  {Y.}~\bibnamefont {Matsui}}, \bibinfo {author} {\bibfnamefont
  {N.}~\bibnamefont {Nagaosa}}, \ and\ \bibinfo {author} {\bibfnamefont
  {Y.}~\bibnamefont {Tokura}},\ }\href {\doibase
  https://doi.org/10.1038/nature09124} {\bibfield  {journal} {\bibinfo
  {journal} {Nature}\ }\textbf {\bibinfo {volume} {465}},\ \bibinfo {pages}
  {901} (\bibinfo {year} {2010})}\BibitemShut {NoStop}%
\bibitem [{\citenamefont {Manton}(1979)}]{Manton:1978gf}%
  \BibitemOpen
  \bibfield  {author} {\bibinfo {author} {\bibfnamefont {N.~S.}\ \bibnamefont
  {Manton}},\ }\href {\doibase 10.1016/0550-3213(79)90309-2} {\bibfield
  {journal} {\bibinfo  {journal} {Nucl. Phys. B}\ }\textbf {\bibinfo {volume}
  {150}},\ \bibinfo {pages} {397} (\bibinfo {year} {1979})}\BibitemShut
  {NoStop}%
\bibitem [{\citenamefont {Taubes}(1980)}]{taubes1980arbitrary}%
  \BibitemOpen
  \bibfield  {author} {\bibinfo {author} {\bibfnamefont {C.~H.}\ \bibnamefont
  {Taubes}},\ }\href {\doibase https://doi.org/10.1007/BF01197552} {\bibfield
  {journal} {\bibinfo  {journal} {Communications in Mathematical Physics}\
  }\textbf {\bibinfo {volume} {72}},\ \bibinfo {pages} {277} (\bibinfo {year}
  {1980})}\BibitemShut {NoStop}%
\bibitem [{\citenamefont {Samols}(1992)}]{Samols:1991ne}%
  \BibitemOpen
  \bibfield  {author} {\bibinfo {author} {\bibfnamefont {T.~M.}\ \bibnamefont
  {Samols}},\ }\href {\doibase 10.1007/BF02099284} {\bibfield  {journal}
  {\bibinfo  {journal} {Commun. Math. Phys.}\ }\textbf {\bibinfo {volume}
  {145}},\ \bibinfo {pages} {149} (\bibinfo {year} {1992})}\BibitemShut
  {NoStop}%
\bibitem [{\citenamefont {Atiyah}\ and\ \citenamefont
  {Hitchin}(2014)}]{atiyah2014geometry}%
  \BibitemOpen
  \bibfield  {author} {\bibinfo {author} {\bibfnamefont {M.~F.}\ \bibnamefont
  {Atiyah}}\ and\ \bibinfo {author} {\bibfnamefont {N.}~\bibnamefont
  {Hitchin}},\ }\href {\doibase https://doi.org/10.1515/9781400859306} {\emph
  {\bibinfo {title} {The geometry and dynamics of magnetic monopoles}}},\
  Vol.~\bibinfo {volume} {11}\ (\bibinfo  {publisher} {Princeton University
  Press},\ \bibinfo {year} {2014})\BibitemShut {NoStop}%
\bibitem [{\citenamefont {Atiyah}\ \emph {et~al.}(1994)\citenamefont {Atiyah},
  \citenamefont {Hitchin}, \citenamefont {Drinfeld},\ and\ \citenamefont
  {Manin}}]{atiyah1994construction}%
  \BibitemOpen
  \bibfield  {author} {\bibinfo {author} {\bibfnamefont {M.~F.}\ \bibnamefont
  {Atiyah}}, \bibinfo {author} {\bibfnamefont {N.~J.}\ \bibnamefont {Hitchin}},
  \bibinfo {author} {\bibfnamefont {V.~G.}\ \bibnamefont {Drinfeld}}, \ and\
  \bibinfo {author} {\bibfnamefont {Y.~I.}\ \bibnamefont {Manin}},\ }\href
  {\doibase https://doi.org/10.1016/0375-9601(78)90141-X} {\bibfield  {journal}
  {\bibinfo  {journal} {Instantons In Gauge Theories}\ ,\ \bibinfo {pages}
  {133}} (\bibinfo {year} {1994})}\BibitemShut {NoStop}%
\bibitem [{\citenamefont {Manton}\ \emph {et~al.}(2021)\citenamefont {Manton},
  \citenamefont {Oles}, \citenamefont {Romanczukiewicz},\ and\ \citenamefont
  {Wereszczynski}}]{Manton:2021ipk}%
  \BibitemOpen
  \bibfield  {author} {\bibinfo {author} {\bibfnamefont {N.~S.}\ \bibnamefont
  {Manton}}, \bibinfo {author} {\bibfnamefont {K.}~\bibnamefont {Oles}},
  \bibinfo {author} {\bibfnamefont {T.}~\bibnamefont {Romanczukiewicz}}, \ and\
  \bibinfo {author} {\bibfnamefont {A.}~\bibnamefont {Wereszczynski}},\ }\href
  {\doibase 10.1103/PhysRevLett.127.071601} {\bibfield  {journal} {\bibinfo
  {journal} {Phys. Rev. Lett.}\ }\textbf {\bibinfo {volume} {127}},\ \bibinfo
  {pages} {071601} (\bibinfo {year} {2021})},\ \Eprint
  {http://arxiv.org/abs/2106.05153} {arXiv:2106.05153 [hep-th]} \BibitemShut
  {NoStop}%
\bibitem [{\citenamefont {Blaschke}\ and\ \citenamefont
  {Karp\'\i{}\v{s}ek}(2022)}]{Blaschke:2022fxp}%
  \BibitemOpen
  \bibfield  {author} {\bibinfo {author} {\bibfnamefont {F.}~\bibnamefont
  {Blaschke}}\ and\ \bibinfo {author} {\bibfnamefont {O.~N.}\ \bibnamefont
  {Karp\'\i{}\v{s}ek}},\ }\href {\doibase 10.1093/ptep/ptac104} {\bibfield
  {journal} {\bibinfo  {journal} {PTEP}\ }\textbf {\bibinfo {volume} {2022}},\
  \bibinfo {pages} {103A01} (\bibinfo {year} {2022})},\ \Eprint
  {http://arxiv.org/abs/2202.05675} {arXiv:2202.05675 [hep-th]} \BibitemShut
  {NoStop}%
\bibitem [{\citenamefont {Adam}\ \emph {et~al.}(2022)\citenamefont {Adam},
  \citenamefont {Manton}, \citenamefont {Oles}, \citenamefont
  {Romanczukiewicz},\ and\ \citenamefont {Wereszczynski}}]{Adam:2021gat}%
  \BibitemOpen
  \bibfield  {author} {\bibinfo {author} {\bibfnamefont {C.}~\bibnamefont
  {Adam}}, \bibinfo {author} {\bibfnamefont {N.~S.}\ \bibnamefont {Manton}},
  \bibinfo {author} {\bibfnamefont {K.}~\bibnamefont {Oles}}, \bibinfo {author}
  {\bibfnamefont {T.}~\bibnamefont {Romanczukiewicz}}, \ and\ \bibinfo {author}
  {\bibfnamefont {A.}~\bibnamefont {Wereszczynski}},\ }\href {\doibase
  10.1103/PhysRevD.105.065012} {\bibfield  {journal} {\bibinfo  {journal}
  {Phys. Rev. D}\ }\textbf {\bibinfo {volume} {105}},\ \bibinfo {pages}
  {065012} (\bibinfo {year} {2022})},\ \Eprint
  {http://arxiv.org/abs/2111.06790} {arXiv:2111.06790 [hep-th]} \BibitemShut
  {NoStop}%
\bibitem [{\citenamefont {Adam}\ \emph {et~al.}(2023)\citenamefont {Adam},
  \citenamefont {Garc\'\i{}a Mart\'\i{}n-Caro}, \citenamefont {Huidobro},
  \citenamefont {Oles}, \citenamefont {Romanczukiewicz},\ and\ \citenamefont
  {Wereszczynski}}]{Adam:2022kla}%
  \BibitemOpen
  \bibfield  {author} {\bibinfo {author} {\bibfnamefont {C.}~\bibnamefont
  {Adam}}, \bibinfo {author} {\bibfnamefont {A.}~\bibnamefont {Garc\'\i{}a
  Mart\'\i{}n-Caro}}, \bibinfo {author} {\bibfnamefont {M.}~\bibnamefont
  {Huidobro}}, \bibinfo {author} {\bibfnamefont {K.}~\bibnamefont {Oles}},
  \bibinfo {author} {\bibfnamefont {T.}~\bibnamefont {Romanczukiewicz}}, \ and\
  \bibinfo {author} {\bibfnamefont {A.}~\bibnamefont {Wereszczynski}},\ }\href
  {\doibase 10.1016/j.physletb.2023.137728} {\bibfield  {journal} {\bibinfo
  {journal} {Phys. Lett. B}\ }\textbf {\bibinfo {volume} {838}},\ \bibinfo
  {pages} {137728} (\bibinfo {year} {2023})},\ \Eprint
  {http://arxiv.org/abs/2212.11936} {arXiv:2212.11936 [hep-th]} \BibitemShut
  {NoStop}%
\bibitem [{\citenamefont {Babaev}\ and\ \citenamefont
  {Speight}(2005)}]{babaev2005semi}%
  \BibitemOpen
  \bibfield  {author} {\bibinfo {author} {\bibfnamefont {E.}~\bibnamefont
  {Babaev}}\ and\ \bibinfo {author} {\bibfnamefont {M.}~\bibnamefont
  {Speight}},\ }\href {\doibase https://doi.org/10.1103/PhysRevB.72.180502}
  {\bibfield  {journal} {\bibinfo  {journal} {Physical Review B}\ }\textbf
  {\bibinfo {volume} {72}},\ \bibinfo {pages} {180502} (\bibinfo {year}
  {2005})}\BibitemShut {NoStop}%
\bibitem [{\citenamefont {Manton}\ and\ \citenamefont
  {Merabet}(1997)}]{Manton:1996ex}%
  \BibitemOpen
  \bibfield  {author} {\bibinfo {author} {\bibfnamefont {N.~S.}\ \bibnamefont
  {Manton}}\ and\ \bibinfo {author} {\bibfnamefont {H.}~\bibnamefont
  {Merabet}},\ }\href {\doibase 10.1088/0951-7715/10/1/002} {\bibfield
  {journal} {\bibinfo  {journal} {Nonlinearity}\ }\textbf {\bibinfo {volume}
  {10}},\ \bibinfo {pages} {3} (\bibinfo {year} {1997})},\ \Eprint
  {http://arxiv.org/abs/hep-th/9605038} {arXiv:hep-th/9605038} \BibitemShut
  {NoStop}%
\bibitem [{\citenamefont {Leese}\ \emph {et~al.}(1995)\citenamefont {Leese},
  \citenamefont {Manton},\ and\ \citenamefont {Schroers}}]{Leese:1994hb}%
  \BibitemOpen
  \bibfield  {author} {\bibinfo {author} {\bibfnamefont {R.~A.}\ \bibnamefont
  {Leese}}, \bibinfo {author} {\bibfnamefont {N.~S.}\ \bibnamefont {Manton}}, \
  and\ \bibinfo {author} {\bibfnamefont {B.~J.}\ \bibnamefont {Schroers}},\
  }\href {\doibase 10.1016/0550-3213(95)00052-6} {\bibfield  {journal}
  {\bibinfo  {journal} {Nucl. Phys. B}\ }\textbf {\bibinfo {volume} {442}},\
  \bibinfo {pages} {228} (\bibinfo {year} {1995})},\ \Eprint
  {http://arxiv.org/abs/hep-ph/9502405} {arXiv:hep-ph/9502405} \BibitemShut
  {NoStop}%
\bibitem [{\citenamefont {Halcrow}(2016)}]{Halcrow:2015rvz}%
  \BibitemOpen
  \bibfield  {author} {\bibinfo {author} {\bibfnamefont {C.~J.}\ \bibnamefont
  {Halcrow}},\ }\href {\doibase 10.1016/j.nuclphysb.2016.01.011} {\bibfield
  {journal} {\bibinfo  {journal} {Nucl. Phys. B}\ }\textbf {\bibinfo {volume}
  {904}},\ \bibinfo {pages} {106} (\bibinfo {year} {2016})},\ \Eprint
  {http://arxiv.org/abs/1511.00682} {arXiv:1511.00682 [hep-th]} \BibitemShut
  {NoStop}%
\bibitem [{\citenamefont {Adam}\ \emph {et~al.}(2010)\citenamefont {Adam},
  \citenamefont {Sanchez-Guillen},\ and\ \citenamefont
  {Wereszczynski}}]{Adam:2010fg}%
  \BibitemOpen
  \bibfield  {author} {\bibinfo {author} {\bibfnamefont {C.}~\bibnamefont
  {Adam}}, \bibinfo {author} {\bibfnamefont {J.}~\bibnamefont
  {Sanchez-Guillen}}, \ and\ \bibinfo {author} {\bibfnamefont {A.}~\bibnamefont
  {Wereszczynski}},\ }\href {\doibase 10.1016/j.physletb.2010.06.025}
  {\bibfield  {journal} {\bibinfo  {journal} {Phys. Lett. B}\ }\textbf
  {\bibinfo {volume} {691}},\ \bibinfo {pages} {105} (\bibinfo {year}
  {2010})},\ \Eprint {http://arxiv.org/abs/1001.4544} {arXiv:1001.4544
  [hep-th]} \BibitemShut {NoStop}%
\bibitem [{\citenamefont {Gillard}\ \emph {et~al.}(2015)\citenamefont
  {Gillard}, \citenamefont {Harland},\ and\ \citenamefont
  {Speight}}]{Gillard:2015eia}%
  \BibitemOpen
  \bibfield  {author} {\bibinfo {author} {\bibfnamefont {M.}~\bibnamefont
  {Gillard}}, \bibinfo {author} {\bibfnamefont {D.}~\bibnamefont {Harland}}, \
  and\ \bibinfo {author} {\bibfnamefont {M.}~\bibnamefont {Speight}},\ }\href
  {\doibase 10.1016/j.nuclphysb.2015.04.005} {\bibfield  {journal} {\bibinfo
  {journal} {Nucl. Phys. B}\ }\textbf {\bibinfo {volume} {895}},\ \bibinfo
  {pages} {272} (\bibinfo {year} {2015})},\ \Eprint
  {http://arxiv.org/abs/1501.05455} {arXiv:1501.05455 [hep-th]} \BibitemShut
  {NoStop}%
\bibitem [{\citenamefont {Naya}\ and\ \citenamefont
  {Sutcliffe}(2018)}]{Naya:2018kyi}%
  \BibitemOpen
  \bibfield  {author} {\bibinfo {author} {\bibfnamefont {C.}~\bibnamefont
  {Naya}}\ and\ \bibinfo {author} {\bibfnamefont {P.}~\bibnamefont
  {Sutcliffe}},\ }\href {\doibase 10.1103/PhysRevLett.121.232002} {\bibfield
  {journal} {\bibinfo  {journal} {Phys. Rev. Lett.}\ }\textbf {\bibinfo
  {volume} {121}},\ \bibinfo {pages} {232002} (\bibinfo {year} {2018})},\
  \Eprint {http://arxiv.org/abs/1811.02064} {arXiv:1811.02064 [hep-th]}
  \BibitemShut {NoStop}%
\bibitem [{\citenamefont {Meier}\ and\ \citenamefont
  {Walliser}(1997)}]{Meier:1996ng}%
  \BibitemOpen
  \bibfield  {author} {\bibinfo {author} {\bibfnamefont {F.}~\bibnamefont
  {Meier}}\ and\ \bibinfo {author} {\bibfnamefont {H.}~\bibnamefont
  {Walliser}},\ }\href {\doibase 10.1016/S0370-1573(97)00012-4} {\bibfield
  {journal} {\bibinfo  {journal} {Phys. Rept.}\ }\textbf {\bibinfo {volume}
  {289}},\ \bibinfo {pages} {383} (\bibinfo {year} {1997})},\ \Eprint
  {http://arxiv.org/abs/hep-ph/9602359} {arXiv:hep-ph/9602359} \BibitemShut
  {NoStop}%
\bibitem [{\citenamefont {Gudnason}\ and\ \citenamefont
  {Halcrow}(2023)}]{Gudnason:2023jpq}%
  \BibitemOpen
  \bibfield  {author} {\bibinfo {author} {\bibfnamefont {S.~B.}\ \bibnamefont
  {Gudnason}}\ and\ \bibinfo {author} {\bibfnamefont {C.}~\bibnamefont
  {Halcrow}},\ }\href@noop {} {\  (\bibinfo {year} {2023})},\ \Eprint
  {http://arxiv.org/abs/2307.09272} {arXiv:2307.09272 [hep-th]} \BibitemShut
  {NoStop}%
\bibitem [{\citenamefont {Halcrow}\ and\ \citenamefont
  {Babaev}(2023)}]{Halcrow:2022kfo}%
  \BibitemOpen
  \bibfield  {author} {\bibinfo {author} {\bibfnamefont {C.}~\bibnamefont
  {Halcrow}}\ and\ \bibinfo {author} {\bibfnamefont {E.}~\bibnamefont
  {Babaev}},\ }\href {\doibase 10.3842/SIGMA.2023.034} {\bibfield  {journal}
  {\bibinfo  {journal} {SIGMA}\ }\textbf {\bibinfo {volume} {19}},\ \bibinfo
  {pages} {034} (\bibinfo {year} {2023})},\ \Eprint
  {http://arxiv.org/abs/2211.02413} {arXiv:2211.02413 [hep-th]} \BibitemShut
  {NoStop}%
\bibitem [{\citenamefont {Portugues}\ and\ \citenamefont
  {Townsend}(2002)}]{portugues2002intersoliton}%
  \BibitemOpen
  \bibfield  {author} {\bibinfo {author} {\bibfnamefont {R.}~\bibnamefont
  {Portugues}}\ and\ \bibinfo {author} {\bibfnamefont {P.~K.}\ \bibnamefont
  {Townsend}},\ }\href@noop {} {\bibfield  {journal} {\bibinfo  {journal}
  {Physics Letters B}\ }\textbf {\bibinfo {volume} {530}},\ \bibinfo {pages}
  {227} (\bibinfo {year} {2002})}\BibitemShut {NoStop}%
\bibitem [{\citenamefont {Alonso-Izquierdo}\ \emph {et~al.}(2021)\citenamefont
  {Alonso-Izquierdo}, \citenamefont {Leon}, \citenamefont {Vaquero},\ and\
  \citenamefont {Mayado}}]{Alonso-Izquierdo:2021tqz}%
  \BibitemOpen
  \bibfield  {author} {\bibinfo {author} {\bibfnamefont {A.}~\bibnamefont
  {Alonso-Izquierdo}}, \bibinfo {author} {\bibfnamefont {M.~A.~G.}\
  \bibnamefont {Leon}}, \bibinfo {author} {\bibfnamefont {J.~M.}\ \bibnamefont
  {Vaquero}}, \ and\ \bibinfo {author} {\bibfnamefont {M.~d. l.~T.}\
  \bibnamefont {Mayado}},\ }\href {\doibase 10.1016/j.cnsns.2021.106011}
  {\bibfield  {journal} {\bibinfo  {journal} {Commun. Nonlinear Sci. Numer.
  Simul.}\ }\textbf {\bibinfo {volume} {103}},\ \bibinfo {pages} {106011}
  (\bibinfo {year} {2021})},\ \Eprint {http://arxiv.org/abs/2105.05750}
  {arXiv:2105.05750 [hep-th]} \BibitemShut {NoStop}%
\bibitem [{\citenamefont {Schroers}(1994)}]{Schroers:1993yk}%
  \BibitemOpen
  \bibfield  {author} {\bibinfo {author} {\bibfnamefont {B.~J.}\ \bibnamefont
  {Schroers}},\ }\href {\doibase 10.1007/BF01413188} {\bibfield  {journal}
  {\bibinfo  {journal} {Z. Phys. C}\ }\textbf {\bibinfo {volume} {61}},\
  \bibinfo {pages} {479} (\bibinfo {year} {1994})},\ \Eprint
  {http://arxiv.org/abs/hep-ph/9308236} {arXiv:hep-ph/9308236} \BibitemShut
  {NoStop}%
\bibitem [{\citenamefont {Feist}(2012)}]{Feist:2011aa}%
  \BibitemOpen
  \bibfield  {author} {\bibinfo {author} {\bibfnamefont {D.~T.~J.}\
  \bibnamefont {Feist}},\ }\href {\doibase 10.1007/JHEP02(2012)100} {\bibfield
  {journal} {\bibinfo  {journal} {JHEP}\ }\textbf {\bibinfo {volume} {02}},\
  \bibinfo {pages} {100} (\bibinfo {year} {2012})},\ \Eprint
  {http://arxiv.org/abs/1112.2119} {arXiv:1112.2119 [hep-th]} \BibitemShut
  {NoStop}%
\bibitem [{\citenamefont {Barton-Singer}\ and\ \citenamefont
  {Schroers}(2023)}]{Barton-Singer:2022rov}%
  \BibitemOpen
  \bibfield  {author} {\bibinfo {author} {\bibfnamefont {B.}~\bibnamefont
  {Barton-Singer}}\ and\ \bibinfo {author} {\bibfnamefont {B.~J.}\ \bibnamefont
  {Schroers}},\ }\href {\doibase 10.21468/SciPostPhys.15.1.011} {\bibfield
  {journal} {\bibinfo  {journal} {SciPost Phys.}\ }\textbf {\bibinfo {volume}
  {15}},\ \bibinfo {pages} {011} (\bibinfo {year} {2023})},\ \Eprint
  {http://arxiv.org/abs/2211.08017} {arXiv:2211.08017 [cond-mat.mes-hall]}
  \BibitemShut {NoStop}%
\bibitem [{\citenamefont {Lund}\ and\ \citenamefont
  {Regge}(1976)}]{Lund:1976ze}%
  \BibitemOpen
  \bibfield  {author} {\bibinfo {author} {\bibfnamefont {F.}~\bibnamefont
  {Lund}}\ and\ \bibinfo {author} {\bibfnamefont {T.}~\bibnamefont {Regge}},\
  }\href {\doibase 10.1103/PhysRevD.14.1524} {\bibfield  {journal} {\bibinfo
  {journal} {Phys. Rev. D}\ }\textbf {\bibinfo {volume} {14}},\ \bibinfo
  {pages} {1524} (\bibinfo {year} {1976})}\BibitemShut {NoStop}%
\bibitem [{\citenamefont {Peyrard}\ and\ \citenamefont
  {Campbell}(1983)}]{Peyrard:1983rzn}%
  \BibitemOpen
  \bibfield  {author} {\bibinfo {author} {\bibfnamefont {M.}~\bibnamefont
  {Peyrard}}\ and\ \bibinfo {author} {\bibfnamefont {D.~K.}\ \bibnamefont
  {Campbell}},\ }\href@noop {} {\bibfield  {journal} {\bibinfo  {journal}
  {Physica D}\ }\textbf {\bibinfo {volume} {9}},\ \bibinfo {pages} {33}
  (\bibinfo {year} {1983})}\BibitemShut {NoStop}%
\bibitem [{\citenamefont {Ferreira}\ and\ \citenamefont
  {Zakrzewski}(2014)}]{Ferreira:2013xda}%
  \BibitemOpen
  \bibfield  {author} {\bibinfo {author} {\bibfnamefont {L.~A.}\ \bibnamefont
  {Ferreira}}\ and\ \bibinfo {author} {\bibfnamefont {W.~J.}\ \bibnamefont
  {Zakrzewski}},\ }\href {\doibase 10.1007/JHEP01(2014)058} {\bibfield
  {journal} {\bibinfo  {journal} {JHEP}\ }\textbf {\bibinfo {volume} {01}},\
  \bibinfo {pages} {058} (\bibinfo {year} {2014})},\ \Eprint
  {http://arxiv.org/abs/1308.4412} {arXiv:1308.4412 [hep-th]} \BibitemShut
  {NoStop}%
\bibitem [{\citenamefont {Navarro-Obreg\'on}\ \emph {et~al.}(2023)\citenamefont
  {Navarro-Obreg\'on}, \citenamefont {Nieto},\ and\ \citenamefont
  {Queiruga}}]{Navarro-Obregon:2023hqe}%
  \BibitemOpen
  \bibfield  {author} {\bibinfo {author} {\bibfnamefont {S.}~\bibnamefont
  {Navarro-Obreg\'on}}, \bibinfo {author} {\bibfnamefont {L.~M.}\ \bibnamefont
  {Nieto}}, \ and\ \bibinfo {author} {\bibfnamefont {J.~M.}\ \bibnamefont
  {Queiruga}},\ }\href {\doibase 10.1103/PhysRevE.108.044216} {\bibfield
  {journal} {\bibinfo  {journal} {Phys. Rev. E}\ }\textbf {\bibinfo {volume}
  {108}},\ \bibinfo {pages} {044216} (\bibinfo {year} {2023})},\ \Eprint
  {http://arxiv.org/abs/2305.00497} {arXiv:2305.00497 [hep-th]} \BibitemShut
  {NoStop}%
\bibitem [{\citenamefont {Cahill}\ \emph {et~al.}(1976)\citenamefont {Cahill},
  \citenamefont {Comtet},\ and\ \citenamefont {Glauber}}]{Cahill:1976im}%
  \BibitemOpen
  \bibfield  {author} {\bibinfo {author} {\bibfnamefont {K.~E.}\ \bibnamefont
  {Cahill}}, \bibinfo {author} {\bibfnamefont {A.}~\bibnamefont {Comtet}}, \
  and\ \bibinfo {author} {\bibfnamefont {R.~J.}\ \bibnamefont {Glauber}},\
  }\href {\doibase 10.1016/0370-2693(76)90202-1} {\bibfield  {journal}
  {\bibinfo  {journal} {Phys. Lett. B}\ }\textbf {\bibinfo {volume} {64}},\
  \bibinfo {pages} {283} (\bibinfo {year} {1976})}\BibitemShut {NoStop}%
\bibitem [{\citenamefont {Walet}(1996)}]{Walet:1996he}%
  \BibitemOpen
  \bibfield  {author} {\bibinfo {author} {\bibfnamefont {N.~R.}\ \bibnamefont
  {Walet}},\ }\href {\doibase 10.1016/0375-9474(96)00219-9} {\bibfield
  {journal} {\bibinfo  {journal} {Nucl. Phys. A}\ }\textbf {\bibinfo {volume}
  {606}},\ \bibinfo {pages} {429} (\bibinfo {year} {1996})},\ \Eprint
  {http://arxiv.org/abs/hep-ph/9603273} {arXiv:hep-ph/9603273} \BibitemShut
  {NoStop}%
\bibitem [{\citenamefont {Walliser}\ and\ \citenamefont
  {Holzwarth}(2000)}]{Walliser:1999ug}%
  \BibitemOpen
  \bibfield  {author} {\bibinfo {author} {\bibfnamefont {H.}~\bibnamefont
  {Walliser}}\ and\ \bibinfo {author} {\bibfnamefont {G.}~\bibnamefont
  {Holzwarth}},\ }\href {\doibase 10.1103/PhysRevB.61.2819} {\bibfield
  {journal} {\bibinfo  {journal} {Phys. Rev. B}\ }\textbf {\bibinfo {volume}
  {61}},\ \bibinfo {pages} {2819} (\bibinfo {year} {2000})},\ \Eprint
  {http://arxiv.org/abs/hep-ph/9907492} {arXiv:hep-ph/9907492} \BibitemShut
  {NoStop}%
\bibitem [{\citenamefont {Evslin}(2021)}]{Evslin:2021vgk}%
  \BibitemOpen
  \bibfield  {author} {\bibinfo {author} {\bibfnamefont {J.}~\bibnamefont
  {Evslin}},\ }\href {\doibase 10.1016/j.physletb.2021.136628} {\bibfield
  {journal} {\bibinfo  {journal} {Phys. Lett. B}\ }\textbf {\bibinfo {volume}
  {822}},\ \bibinfo {pages} {136628} (\bibinfo {year} {2021})},\ \Eprint
  {http://arxiv.org/abs/2109.05852} {arXiv:2109.05852 [hep-th]} \BibitemShut
  {NoStop}%
\end{thebibliography}%
\bibliographystyle{apsrev4-1}

\end{document}